\documentclass[fleqn,usenatbib]{mnras}

\usepackage{txfonts}
\usepackage[T1]{fontenc}
\usepackage{ae,aecompl}
\usepackage{array}

\usepackage{graphicx}	% Including figure files
\usepackage[T1]{fontenc}
\usepackage[normalem]{ulem}
\newcommand{\cd}{d$^{-1}$\,}
\title[Search for exoplanets around pulsating stars and the case of KIC 8197761]{Search for exoplanets around pulsating stars of A--F type in \textit{Kepler} Short Cadence data and the case of KIC 8197761}

\author[Paulina Sowicka et al.]{Paulina 
Sowicka,$^{1,2}$\thanks{E-mail:paula@camk.edu.pl} Gerald Handler,$^{2}$ Bart\l{}omiej D\k{e}bski,$^{1}$ David Jones,$^{3,4}$ \and Marie Van de Sande,$^5$ and P{\'e}ter I. P{\'a}pics\,$^5$\\
$^{1}$Astronomical Observatory of the Jagiellonian University, Orla 171, Cracow, Poland\\
$^{2}$Nicolaus Copernicus Astronomical Centre, Bartycka 18, Warsaw, Poland\\
$^{3}$Instituto de Astrof\'isica de Canarias, E-38205 La Laguna, Tenerife, Spain\\
$^{4}$Departamento de Astrof\'isica, Universidad de La Laguna, E-38206 La Laguna, Tenerife, Spain\\
$^{5}$Instituut voor Sterrenkunde, KU Leuven, Celestijnenlaan 200D, B-3001 Leuven, Belgium}

\date{Accepted XXX. Received YYY; in original form ZZZ}

\pubyear{2016}

\begin{document}
\label{firstpage}
\pagerange{\pageref{firstpage}--\pageref{lastpage}}
\maketitle

\begin{abstract}
We searched for extrasolar planets around pulsating stars by examining 
\textit{Kepler} data for transit-like events hidden in the intrinsic variability. 
All Short Cadence observations for targets with 6000~K 
$< T_{\rm eff} <$ 8500~K 
were visually inspected for transit-like events following the removal of 
pulsational signals by sinusoidal fits. Clear transit-like events were 
detected in KIC 5613330 and KIC 8197761. KIC 5613330 is a confirmed 
exoplanet host (Kepler-635b), where the transit period determined here 
is consistent with the literature value. KIC 8197761 is a $\gamma$ Doradus - $\delta$ 
Scuti star exhibiting eclipses/transits occurring every 9.8686667(27) d, 
having durations of 8.37~h, and causing brightness drops $\frac{\Delta F}{F} = 
0.00629(29)$. The star's pulsation spectrum contains several mode 
doublets and triplets, identified as $l = 1$, with a mean spacing of 
0.001659(15) \cd, implying an internal rotation period of $301\pm3$ d. 
Trials to calculate the size of the light travel time effect (LTTE) from 
the pulsations to constrain the companion's mass ended inconclusive. 
Finding planets around $\gamma$ Doradus stars from the pulsational LTTE, 
therefore, is concluded to be unrealistic. Spectroscopic monitoring of KIC 
8197761 revealed sinusoidal radial velocity variations with a 
semi-amplitude of $19.75 \pm 0.32$ km/s, while individual spectra present 
rotational broadening consistent with $v \sin i~=~9\pm1$ km/s. This 
suggests that the stellar surface rotation is synchronized with the 
orbit, whereas the stellar core rotates $\sim$30 times slower. Combining the 
observed radial velocity variability with the transit photometry, 
constrains the companion's mass to be $\approx 0.28$~M$_{\odot}$, ruling out an exoplanet 
hypothesis.
\end{abstract}


%%%%%%%%%%%%%%%%%%%%%%%%%%%%%%%%%%%%%%%%%%%%%%%%%%

%%%%%%%%%%%%%%%%% BODY OF PAPER %%%%%%%%%%%%%%%%%%

\section{Introduction}

The observational study of both transiting exoplanets and stellar pulsations has experienced enormous progress during the last few years.
Both require 
photometric measurements of the highest possible precision and with high duty cycle. 
Ground-based observations are intrinsically limited in these respects due to the 
influence of the Earth's atmosphere and the day/night cycle. 
Consequently, the deployment of dedicated space photometry missions, in 
particular Kepler (e.g., \citealt{koch2010}), represented a great leap for both fields.

The first discovery of a planet around a star other than our Sun was made 
by \citet{wolszczan1992} through the detection of timing variations, caused by the light travel time effect (LTTE), in the millisecond pulsar PSR1257+12.  Since then, a variety of other methods have been employed to discover many more extrasolar planets, among them the radial velocity method \citep[e.g.\ HARPS -- high precision \'echelle spectrograph;][]{mayor2003} and gravitational microlensing \citep[e.g.\ The OGLE project,][]{udalski1992}.  The most efficient method, however, has proved to be the transit method, with several dedicated wide-field survey instruments having been built to search for exoplanet transits (e.g.\ Super-WASP, \citealt{pollacco2006}; TrES, \citealt{odonovan2006}; HAT, \citealt{bakos2004}; and many more).  The main issue for such photometric surveys is the detection limit -- limited not only by the collecting area of the telescopes employed but also by the Earth's atmosphere. Atmospheric disturbances can be corrected with adaptive optics, but such instrumentation is rarely used for high-precision 
photometry. Therefore astronomers studying exoplanets had to wait for the launch of CoRoT in 2006 \citep{baglin2003} and \textit{Kepler} in 2009 \citep{koch2010} to circumvent this problem.  Both were specifically designed to find planets much smaller than those within the detection limits of ground-based surveys.  Since then, the ``\textit{Space} revolution'' of the field began.

Most stars are variable -- this has been known for decades. The origins of this variability can differ greatly from star to star, but they can be divided into two main groups: extrinsic and intrinsic variables. 
Extrinsic are those whose observed brightness does not change because of physical 
changes of the star, but rather because of other external effects, for example being eclipsed by a binary companion. Intrinsic variables, on the other hand, change their 
physical properties -- like pulsating stars changing their sizes and 
shapes. Given the wide-range of causes for their observed variability, variable stars present a very broad range of amplitudes and periodicities. For example, only considering pulsating stars, the longest 
periods (100 -- 1000 d) are Mira type variables, whilst the shortest (1 
-- 2 min) are pulsating subdwarf O type stars (see, e.g., the summary by \citealt{handler2013}).

Asteroseismology uses stellar oscillations as seismic waves to determine 
the internal structure of stars. It applies the same method as seismology 
to reveal the Earth's interior. By comparing theoretical predictions of 
frequencies of oscillations with observed ones, we can learn about the 
stellar interiors, as the theoretical models must be altered to reproduce 
the observations.

The photometric amplitude of stellar pulsations may be of the order of 
one part per million (the present observational detection level) up to 
several magnitudes. In comparison, the decrease in flux from a transit 
of an Earth-like planet is of the order of $0.01\%$. Transits, whose 
amplitudes are bigger than the amplitudes of the intrinsic brightness variations of 
their host stars are relatively easy to spot. Difficulties arise when 
intrinsic variability begins to exceed that caused by planetary transits, where it 
can be said that such transits are ``hidden'' in pulsations.  However, as transits result in clearly nonsinusoidal lightcurves, harmonic functions are well-suited for the process of detecting and subtracting pulsational signals, allowing the residuals to be searched for planetary transits.  In this work, we search for such planetary transits, which are of particular interest given that the pulsation properties can then be used to learn more about the host stars and hence also their transiting planets.  One such case is that of Kepler-444 \citep{campante2015}, where the 
authors used asteroseismic methods to measure an age of 11 Gyr for the 
star, or Kepler-56 \citep{huber2013}, where asteroseismology was used to 
show that the planetary orbits are misaligned with respect to the 
stellar spin axis.
%-------------------------------------------------------------------------------------------------------------------------
\section[]{Pre-search preparation of data}
Two types of data are available from 
\textit{Kepler}: Long Cadence (LC) and Short Cadence (SC). For Long 
Cadence data 270 frames and for Short Cadence data 9 frames are co-added 
(each frame is 6.02 s exposure time plus 0.52 s readout time). This 
results in a total of 29.4 min and 58.85 s for Long and Short Cadence, 
respectively. Files including Long Cadence data (images and light 
curves) span a quarter, whilst Short Cadence data span a month. There 
are $\sim$156 000 LC and 512 SC targets for each observing interval 
(i.e. quarter for LC, month for SC). Long Cadence data are primarily 
used for planet detection, while Short Cadence data are especially important 
 for asteroseismology, better transit timing \citep{koch2010} 
and transit light curve modelling. Raw pixel data are transferred 
from the spacecraft once per month. Each data set is then processed in parallel 
by the Science Pipeline \citep{jenkins2010}. The resulting raw 
and calibrated light curves with estimates of uncertainties are 
then made available to the public via NASA's Mikulski Archive for Space 
Telescopes (MAST)\footnote{https://archive.stsci.edu/}.

The search criteria used to retrieve data for the search performed in 
this work were as follows:
\begin{enumerate}
\item {Effective temperature} \\
The search was conducted for temperatures in the range between 6000 and 8500 K. Effective temperatures in \textit{Kepler} Input Catalog are from \citet{brown2011} and are accurate to $\sim$200~K.
Such a wide range was needed to include stars of spectral type A and F in the classical instability strip, among which there are many pulsating stars of e.g. $\delta$ Scuti or $\gamma$ Doradus type. \\
\item {Target Type}\\
Only stars with SC data were chosen, because as the cadence of LC data is unsuitable for detailed asteroseismic analysis and revealing the shape of any ``hidden'' transits. 
\end{enumerate}

The search resulted in 7546 individual FITS files, for a total of 2292 
stars. Each FITS file contains multiple columns, from which two were 
taken for the further work:\\ TIME [d] -- Barycentric Kepler Julian Day 
(BKJD: BJD - 2454833) calculated for the target in the file,\\ PDCSAP\_FLUX [e-/s] -- 
Aperture photometry flux after Pre-search Data Conditioning, prepared 
for planet transit search. \\ These two columns were extracted from the 
FITS file and flux values were converted from e-/s to magnitudes, 
using Astropy{\footnote{http://www.astropy.org/, \citet{astropy2013}} routines. All of the light curves were then cleaned 
using a dedicated program which runs a moving-average filter over the 
data and rejects the most obvious single outliers. The process was 
visually supervised to avoid the loss of some possible short planetary 
transits, as those would make themselves obvious in SC data as groups of 
low points.

%------------------------------------------------------------------------------------------------------------------------- 
\section{The procedure}

The cleaned SC data were Fourier searched for the pulsational signals with highest amplitudes, which were then fit and subtracted from the light curve such that the prewhitened light curve could be inspected for transit-like events.  This process was performed automatically and stopped either when no signal in the periodogram exceeded $S/N=4$ or when 60 frequencies had been
prewhitened.  At this point it was not important whether these 
signals were physical or not - the key was to remove the intrinsic 
variability to an extent that makes a search for transit-like signals 
possible. 

The search for transit signals was carried after each round of pulsation signal subtraction. Initially, the Fourier analysis 
and pre-whitening were performed for the full frequency range from 0 to 
80 \cd. Whenever this procedure did not yield satisfactory results, e.g.\ 
in case of low- and high-frequency signals of similar amplitude, two 
separate frequency ranges (below 8 \cd and from 8 to 80 \cd) were 
sequentially analysed. Even so, it was sometimes not possible to remove 
all the intrinsic variations, mostly because of unresolved close 
frequencies that resulted in artificially enhanced noise in the 
periodogram. In any case, this approach was usually sufficient to remove the 
dominant components of the intrinsic variability and facilitate the search for 
transit-like signals. A possible ``comb'' of 
harmonic frequencies caused by transit-like signals in the periodogram 
would be interpreted by this program as an increase in noise level 
rather than individual oscillations and the automatic prewhitening would 
leave such structures unaffected.

The procedure for the transit search itself was inspired by the Planet 
Hunters project{\footnote{planethunters.org}}, i.e. 
all the light curves were first inspected visually. A preliminary pass, where the only selection criterion was 
the occurrence of dips of any kind (not necessarily recurring), resulted in a 
list of 230 candidate stars. After a more careful inspection of those candidates, the list shrank to 81. During that pass, obvious candidates with dips caused by 
instrumental effects were rejected (e.g. dips occurring at the exactly 
same time during one quarter for separate stars). For the last sort, all 
the light curves were critically viewed and this resulted in 42 final 
candidates, listed in Table~\ref{table:candidates}.  It is important to note that, in addition to this 42 candidates, a further 100 stars already marked as ``Planetary candidates'' in the MAST database were also recovered (these stars are not listed here). A comparison of the list of our candidates and already known candidates in MAST with the catalogue from \citet{batalha2013} shows that we recovered 106 candidates out of 130 which also appeared in \citet{batalha2013}.

Of the final 42 candidates, only one showed periodic transit-like events.  In some cases, the dips were not of a typical transit-like profile (with a flat minimum), however, as 
transiting systems exist whose geometry does not generate flat minima in the 
light curves, we decided to not exclude those from our list. There 
are at least a few effects that can change the transit shape, e.g. 
asymmetric transit shapes due to the effects of stellar gravity 
darkening of a rapidly rotating, oblate host star \citep{barnes2009}. 
We caution that some of these events may not be real transits, but would 
rather be due to imperfectly removed pulsations mimicking a transit-like 
light curve shape. This, in turn, may offer an explanation for the candidates where we detect only single events.  As such, we do not claim that all candidates here show true transits, but rather that they are worthy of more in-depth study.

For each of our 42 candidates, Long Cadence data was also inspected. 
We normalised and merged the data, and calculated Fourier spectra 
to search for harmonic frequencies of possible orbital frequencies. As 
transits generate nonsinusoidal variability, a Fourier spectrum of such 
a light curve is expected to contain many harmonic frequencies with 
separations equal to the frequency of repeated transits. Measurement of 
the separation of these harmonics gives an estimation of the 
corresponding period.

Among the analysed candidates, only two showed clearly visible harmonics: 
KIC 5613330 (Fig.~\ref{fig:561pow_spec}) and KIC 8197761 (after 
prewhitening, Fig.~\ref{fig:power_spectrum}). In the case of KIC 5613330, 
the frequency of the transits determined using harmonics is $0.04264(11)$ 
\cd (period of $23.452$ d). The light curve phased at this frequency is 
shown in Fig.~\ref{fig:516transit}. This planet candidate has been 
identified by the {\it Kepler} pipeline as KOI-649.01 with a period very
 close to our determination and later on validated as Kepler-635b by 
\citealt[][with a period of $23.44971(8)$ d]{morton2016}. However, as 
KIC 5613330 did not show pulsational variability that could be used for 
further asteroseismological investigation, we focus on KIC 8197761.

\begin{figure}
  \centering
  \includegraphics[scale=0.41]{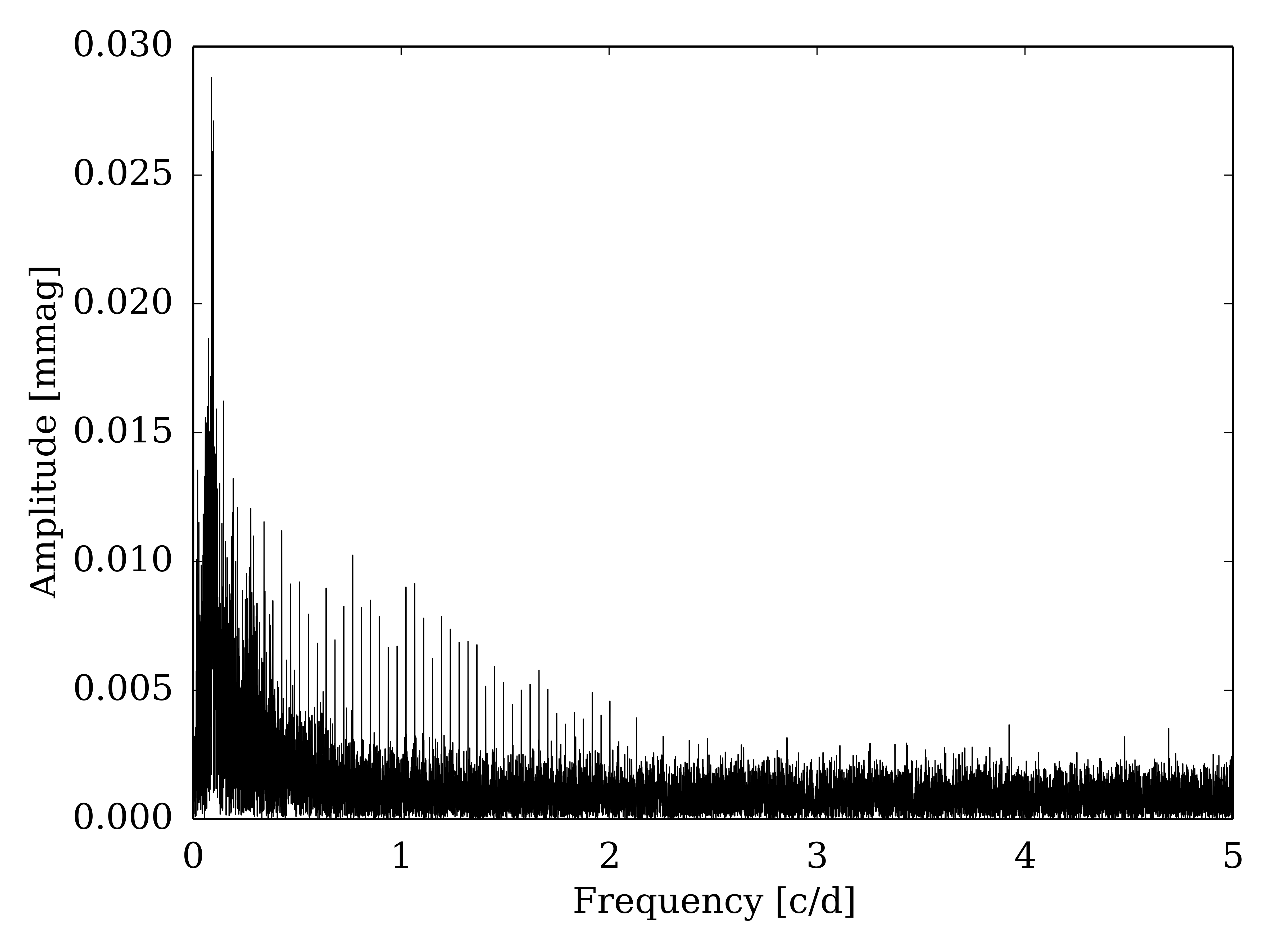}
  \caption[The periodogram of the \textit{Kepler} long cadence observations of KIC 5613330]{The periodogram of the \textit{Kepler} long cadence observations of KIC 5613330 showing a series of harmonic frequencies.}
  \label{fig:561pow_spec}
\end{figure}
%------------
\begin{figure*}
  \centering
  \includegraphics[scale=0.36]{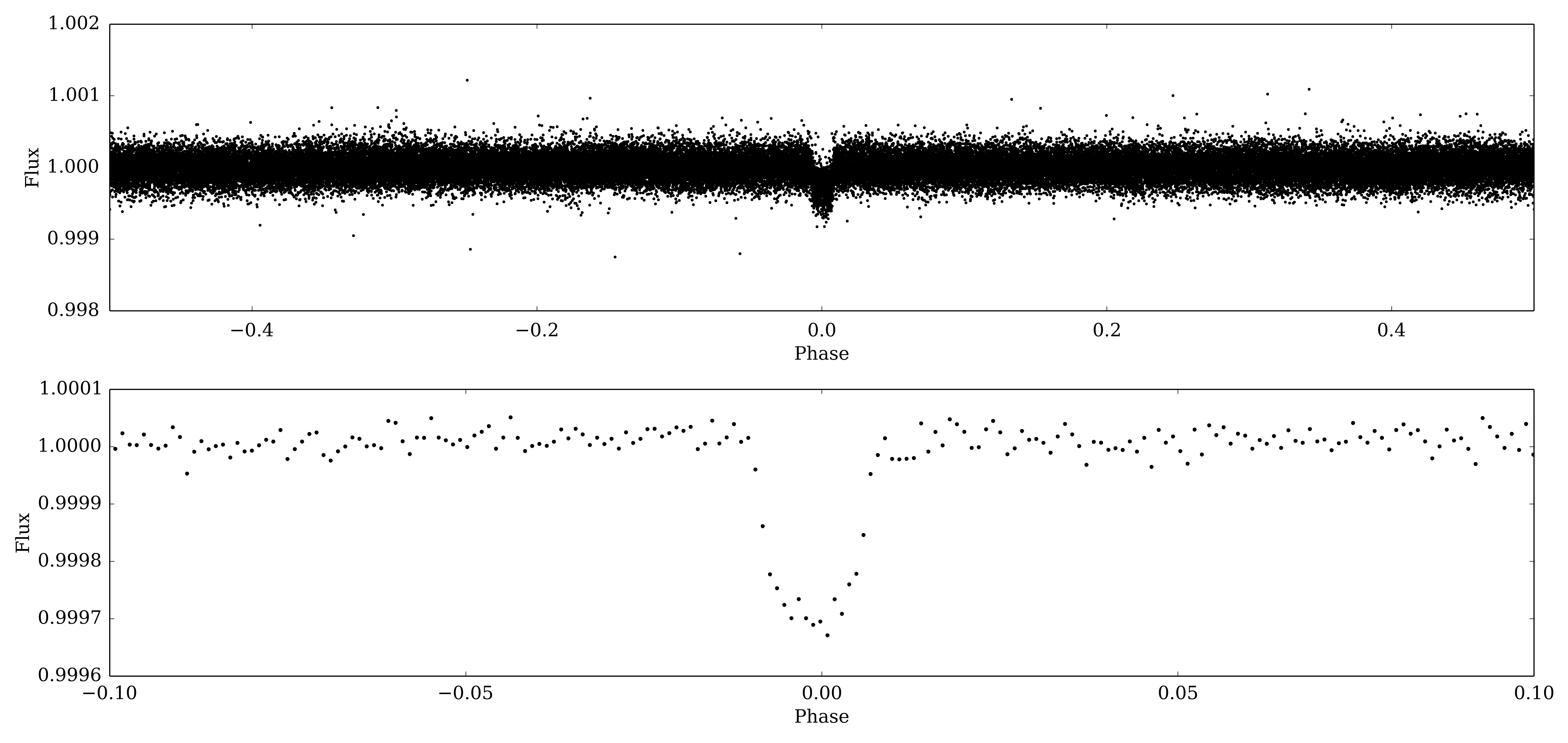}
  \caption[The transit profile of KIC 5613330 in zoom]{Upper panel: The light curve of KIC 5613330 long cadence quarters Q0-Q17 phased at the transit period. Lower panel: The transit profile of KIC 5613330 in zoom. The data is binned in phase with 0.001 bin size.}
  \label{fig:516transit}
\end{figure*}
%-----------
\begin{table*}
\centering
\caption[The list of planetary candidates]{The resulting list of candidates. Each star has noted: the quarter Q of 
observations, in which a transit-like event was spotted, the approximate 
time of the event (in few cases there was more than one event) and the 
type of intrinsic variability (if any). Notes: SLO -- solar-like 
oscillations, $\delta$ Sct -- $\delta$ Scuti-type pulsations, $\gamma$~Dor -- $\gamma$ Doradus-type pulsations, hybrid -- hybrid $\delta$ Scuti- 
and $\gamma$ Doradus-type pulsations. References: $1$ - 
\citet{molenda2013}, $2$ - \citet{uytterhoeven2011}. When no 
reference to the variability type is given, the classification was performed 
by us.}
%{\tiny{
\begin{tabular}{l c c c||l c c c}
\hline
KIC    & Q & Approx. transit time & Variability type                          & KIC     & Q & Approx. transit time & Variability type \\ 
 & & [BJD - 2454833] &                          &  &  & [BJD - 2454833] & \\ \hline
$ 1435467 $   & $ 2 $   & $ 250.5-251.5 $     & SLO$^1$                       &          $ 7976303 $	& $ 10 $  & $ 989.0-989.25 $    & SLO$^1$ 	      \\ 
              & $ 6 $   & $ 573.5-574.0 $     &                               &          $ 8179536 $    & $ 10 $  & $ 935.8-936.0 $     & SLO$^1$	      \\     
$ 1718222 $   & $ 2 $   & $ 180.5-180.6 $     & ?                             &          $ 8197761 $    & $ 1 $	  & $ 138.9-139.1 $     & $\gamma$ Dor$^2$    \\
              & $ 2 $   & $ 185.65-185.75 $   &                               &                         & $ 1 $	  & $ 148.8-148.9 $	&                     \\
$ 1724961 $   & $ 0 $   & $ 128.8-128.85 $    & hybrid                        &                         & $ 1 $	  & $ 158.7-158.8 $	&                     \\
$ 2011424 $   & $ 3 $   & $ 288.55-289.0 $    & hybrid                        &          $ 8525286 $    & $ 2 $	  & $ 185.65-185.75 $   & $\delta$ Sct$^2$    \\
$ 2166218 $   & $ 2 $   & $ 233.3-233.5 $     & $\gamma$ Dor$^2$              &          $ 8694723 $    & $ 10 $  & $ 909.5-910.5 $     & SLO$^1$             \\
$ 2584426 $   & $ 2 $	& $ 208.25-208.75 $   & ?	                      &                         & $ 14 $  & $ 1280.2-1280.6 $   &                     \\ 
$ 3456181 $   & $ 9 $	& $ 822.0-822.5 $     & SLO$^1$	                      &          $ 9139163 $	& $ 2 $	  & $ 222.0-222.3 $     & SLO$^1$	      \\ 
	      & $ 9 $	& $ 840.0-840.25 $    & 	                      & 	                & $ 5 $	  & $ 490.0-490.5 $     &	              \\ 
$ 3837815 $   & $ 3 $	& $ 293.0-293.5 $     & rotation	              &                         & $ 5 $	  & $ 499.5-501.0 $     &	              \\ 
$ 3942392 $   & $ 2 $	& $ 248.2-248.3 $     & rotation                      &          $ 9206432 $	& $ 8 $	  & $ 775.75-777.0 $    & SLO$^1$	      \\ 
$ 4269337 $   & $ 2 $	& $ 241.6-242.0 $     & $\delta$ Sct$^2$              &          $ 9353712 $	& $ 5 $	  & $ 520.0-520.5 $     & ?	              \\ 
$ 4931390 $   & $ 9 $	& $ 891.5-892.0 $     & rotation + $\gamma$ Dor       &                         & $ 11 $  & $ 1038.75-1039.25 $ &                     \\
$ 5024454 $   & $ 9 $	& $ 865.35-865.4 $    & ?                             &          $ 9579208 $	& $ 15 $  & $ 1385.0-1385.25 $  & ?	              \\ 
$ 5200689 $   & $ 11 $  & $ 1008.4-1008.5 $   & constant                      &          $ 9700855 $	& $ 2 $	  & $ 207.0-207.5 $     & ?	              \\ 
$ 5299306 $   & $ 3 $   & $ 269.0-270.0 $     & ?                             &          $ 10010623 $   & $ 2 $   & $ 196.5-196.75 $    & ?                   \\
$ 5613330 $   & $ 8 $	& $ 745.5-746.0 $     & constant    	              &          $ 10068307 $   & $ 16 $  & $ 1493.5-1494.5 $   & SLO$^1$             \\ 
$ 5641711 $   & $ 2 $	& $ 243.75-244.0 $    & hybrid$^2$	              &                         & $ 16 $  & $ 1548.0-1550.0 $   &	              \\ 
$ 5773345 $   & $ 11 $  & $ 1082.0-1082.25 $  & SLO$^1$	                      &          $ 10208303 $   & $ 4 $	  & $ 373.0-374.25 $    & SLO$^2$	      \\
$ 5961597 $   & $ 3 $	& $ 306.25-306.75 $   & variable	              &          $ 10294220 $   & $ 4 $	  & $ 356.5-357.5 $     & rotation            \\
$ 5980337 $   & $ 2 $	& $ 176.5-177.0 $     & $\delta$ Sct, instrumental?   &                         & $ 4 $	  & $ 377.5-378.5 $	&                     \\
$ 6126855 $   & $ 3 $	& $ 300.0-300.5 $     & rotation?	              &          $ 10355856 $   & $ 6 $	  & $ 590.25-590.75 $   & SLO$^1$ + rotation  \\
$ 6508366 $   & $ 9 $	& $ 900.75-901.0 $    & SLO$^1$	                      &          $ 11081729 $	& $ 15 $  & $ 1424.75-1426.0 $  & SLO$^1$	      \\ 
$ 7103006 $   & $ 11 $  & $ 1096.75-1097.0 $  & SLO$^1$                       &          $ 11874676 $	& $ 7 $	  & $ 675.25-675.75 $   & hybrid	      \\ 
$ 7206837 $   & $ 9 $	& $ 850.9-851.0 $     & SLO$^1$/rotation	      &          $ 12647070 $	& $ 4 $	  & $ 372.5-373.5 $     & $\delta$ Sct$^2$    \\    
$ 7548479 $   & $ 10 $  & $ 923.5-924.5 $     & $\delta$ Sct$^2$	      &                         &         &                     &                     \\ \hline 
\end{tabular} 
%}}
\label{table:candidates}
\end{table*} 

%-------------------------------------------------------------------------------------------------------------------------
\section{Analysis of KIC 8197761}
%-------------------------------------------------------------------------------------------------------------------------
\subsection{About KIC 8197761}

\begin{table}
\centering
\caption[KIC 8197761 basic parameters]{KIC 8197761 basic parameters derived from photometry. References: 1 -- \citet{brown2011}. 2 -- Those values were derived from the Dartmouth Stellar Evolution Program Models {\bf by \citet{huber2014}}.}
\begin{tabular}{l c c}
\hline \noalign{\smallskip}
\multicolumn{3}{c}{Basic data} \\ \noalign{\smallskip}
RA (J$2000.0$) & $20^{h}$ $04^{m}$ $09^{s}.3050$ & \\ \noalign{\smallskip}
Dec (J$2000.0$) & $+44^{\circ}$ $04'$ $15.983"$ & \\ \noalign{\smallskip}
Brightness & $m_{Kep}$ = $10.656$ mag & \\ \hline \noalign{\smallskip}
\multicolumn{3}{c}{Stellar properties}	\\ \noalign{\smallskip}
Parameter	& Value	& Reference	 \\ \hline \noalign{\smallskip}
$T_{eff}$ [K]	& $ 7301^{+228}_{-325} $	& $ 1 $	 \\  \noalign{\smallskip}
$\log(g)$ [cm/s$^2$] & $ 4.110^{+0.159}_{-0.286} $	& $ 1 $	 \\ \noalign{\smallskip}
[Fe/H] [Sun]	& $ -0.380^{+0.240}_{-0.360} $	& $ 1 $	 \\ \noalign{\smallskip}
$R$ [$R_{\sun}$] 	& $ 1.717^{+0.858}_{-0.410} $	& $ 2 $	 \\ \noalign{\smallskip}
$M$ [$M_{\sun}$] 	& $ 1.384^{+0.281}_{-0.276} $	& $ 2 $	 \\ \noalign{\smallskip}
$\rho$ [g/cm$^3$] & $ 0.3850^{+0.3367}_{-0.2517} $	& $ 2 $	 \\ \noalign{\smallskip} \hline 
\end{tabular}
\label{table:parameters}
\end{table}
Basic information about KIC 8197761 is presented in 
Table~\ref{table:parameters}.  The star was observed in LC in all quarters (Q0-Q17), whilst in SC only 
in Q1. It is located in the field of the open cluster NGC 6866, but 
according to \citet{kharchenko2004}, \citet{molenda2009}, 
\citet{frolov2010} and \citet{bost2015} it is not a cluster member 
(membership probability equal to zero). A literature search revealed that 
contradictory variability types have been reported. The star was classified as 
$\gamma$ Doradus type by \citet{uytterhoeven2011} with no comments on 
additional features like transit-like events. \citet{slawson2011} 
published a catalog of eclipsing binary stars after the second 
\textit{Kepler} data release and in this catalog KIC 8197761 was 
classified as a detached binary (Algol-type) with a period of 
$19.738450$ days, ratio of secondary to primary temperature $T2/T1=0.770$, scaled sum of radii $R1+R2=0.129$ and $\sin i=0.99497$, but 
without any estimation of masses. \citet{tenenbaum2012} 
published a list of detections of potential transit signals and an entry 
for KIC 8197761 was present -- with a period of $9.87$ days and transit 
depth $3030.9$ in parts per million. A priori, it is unclear which of these classifications is correct.  Here, we attempt to resolve this issue under the basic assumption that the observed transits are associated with KIC8197761, and not due to a background eclipsing binary 
accidentally falling into the photometric aperture. The estimated 
contamination factor in MAST is 1.2\%, low enough that the possible contamination was ignored in our analysis of the light curve.

%-------------------------------------------------------------------------------------------------------------------------
\subsection{Determining the observables from the light curve}
\label{sec:per}
Due to the fact that KIC 8197761 was observed in SC only in Quarter 1 
and there were only three transit-like features visible, it was decided 
to include the observations in LC for all the quarters, Q0-Q17. Raw FITS 
files were downloaded from the MAST archive and each of them was corrected 
using \textit{Kepler} cotrending basis vectors via 
\texttt{PyKE}\footnote{http://keplerscience.arc.nasa.gov/PyKE.shtml}. 
Then all the light curves were merged together and normalised in flux; 
the resulting light curve is shown in Fig.~\ref{fig:detrended}. 
\begin{figure*}
  \centering
  \includegraphics[scale=0.36]{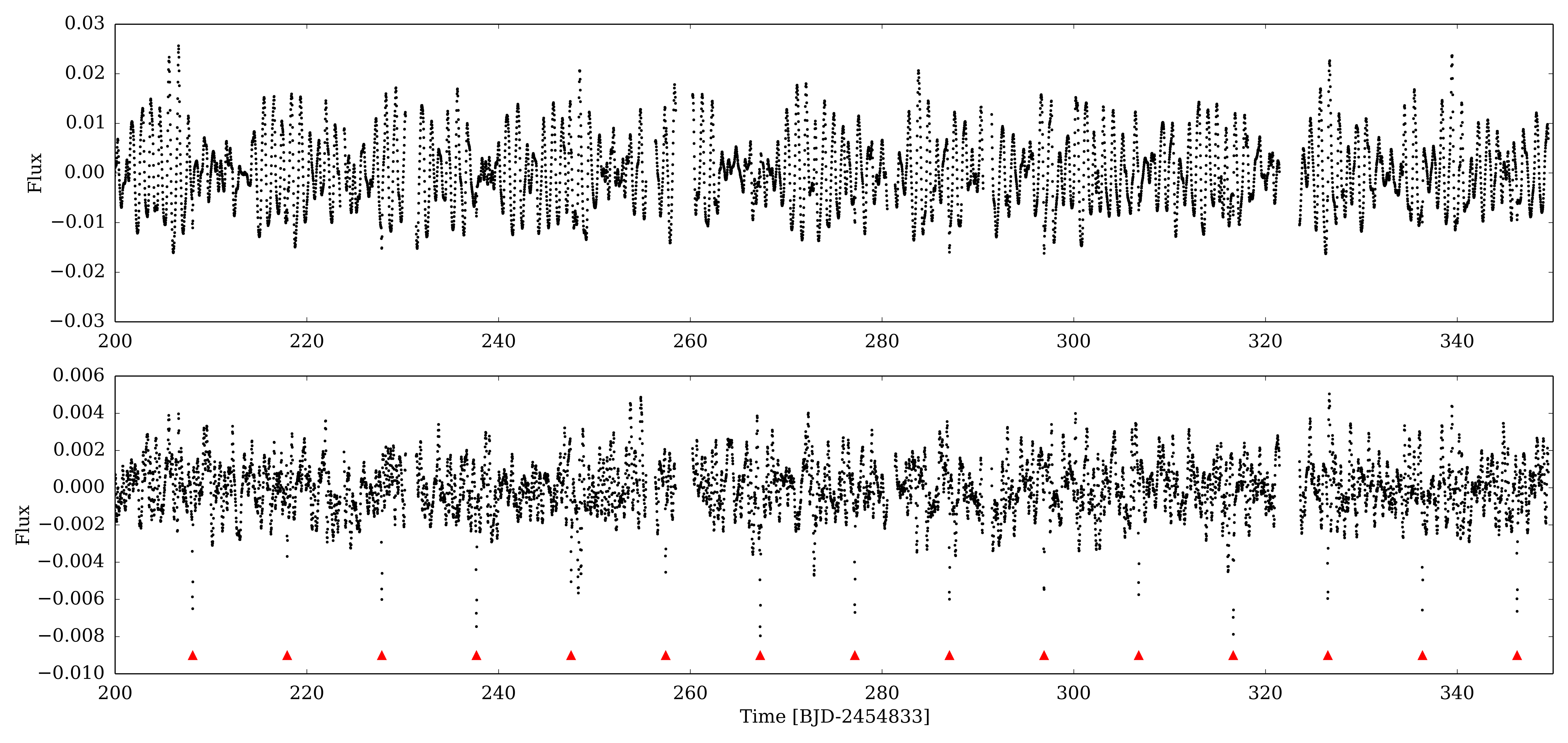}
  \caption[Long cadence \textit{Kepler} observations of KIC 8197761 after detrending and normalisation]{The upper panel shows a portion of long cadence (days 200-350 from the start of the mission, approx. Q4-Q5) \textit{Kepler} observations of KIC 8197761 after detrending, 
  normalizing in flux, and merging. Multi-periodic $\gamma$~Doradus pulsations are clearly visible, while transit-like events are not visible at all. The lower panel shows the same portion of data after removal of 53 pulsation frequencies. Transit-like events are now visible and are indicated by triangles. Note the different ordinate scale of the upper and lower panel.}
  \label{fig:detrended}
\end{figure*}
\begin{figure}
  \centering
  \includegraphics[scale=0.41]{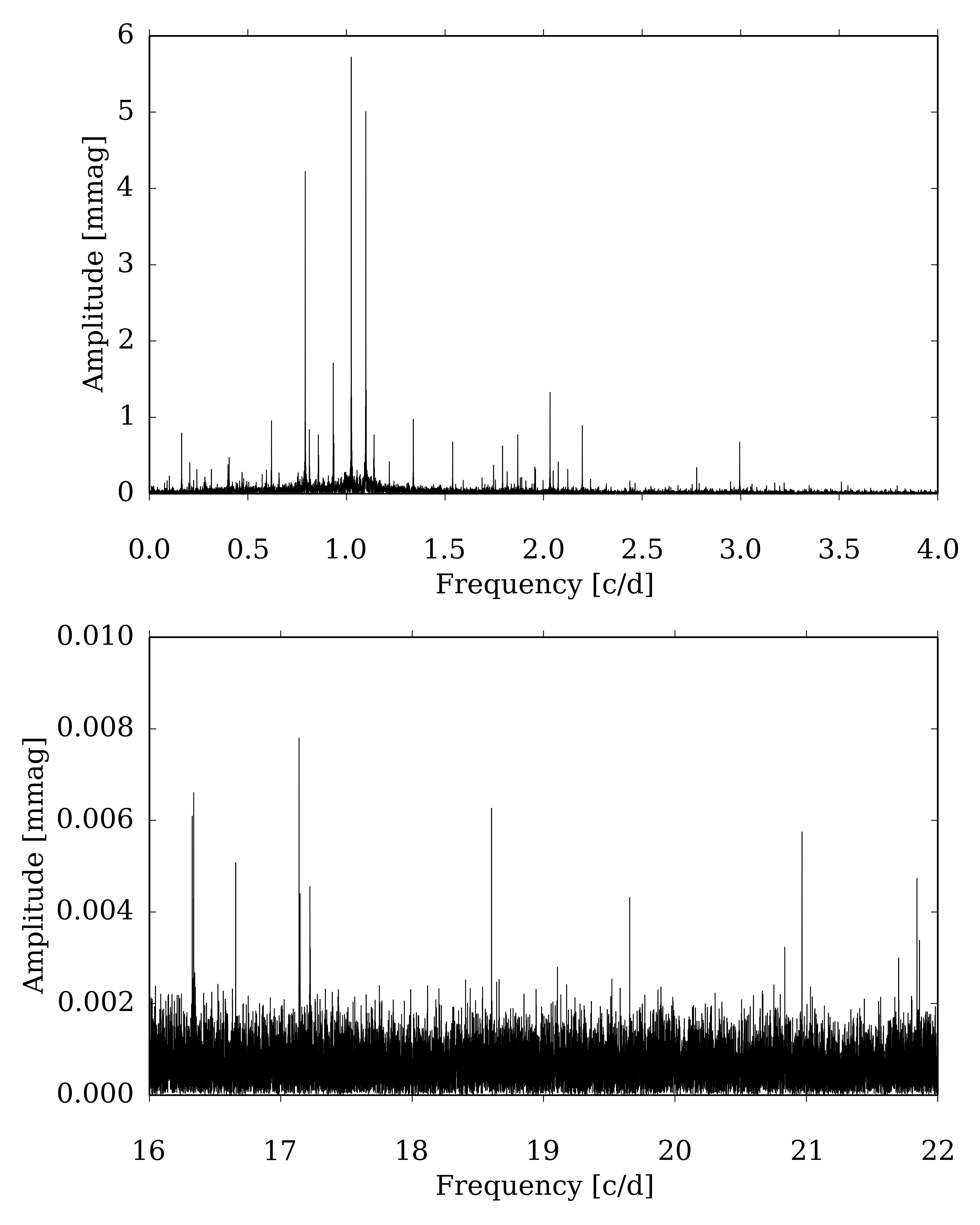}

  \caption[Upper panel: the periodogram of the \textit{Kepler} long 
cadence observations of KIC 8197761]{Upper panel: the periodogram of the 
\textit{Kepler} long cadence observations of KIC 8197761. Lower panel: 
periodogram in the high-frequency region after prewhitening the 
low-frequency oscillations. Some low-amplitude $\delta$ Scuti pulsations 
can be discerned.}
  \label{fig:power_spectrum}
\end{figure}

The next step was to determine the variability content of the light 
curve. We used the LC data for this purpose to take advantage of their 
longer time base. However, the crude procedure previously employed in order to facilitate the 
detection of transit-like signals is no longer sufficent. Instead, we 
used the program \texttt{Period04}\footnote{which was also used for all 
the further calculations} \citep{lenz2005} to perform an 
interactive frequency analysis, on the out-of-eclipse data only.

The Fourier spectrum is shown in Fig.~\ref{fig:power_spectrum}. Several 
hundreds of frequencies are present in the frequency range from 0.1 to 5 
\cd (Fig.~\ref{fig:power_spectrum}, upper panel). To avoid over-interpretation of the data, we only accepted frequencies 
exceeding an arbitrary amplitude limit of 0.17 mmag. The frequencies and 
amplitudes of those 65 signals are listed in 
Table~\ref{table:sc_frequencies}.

\begin{table*}
\centering
\caption{The highest-amplitude periodic signals detected in the 
\textit{Kepler} long cadence data of KIC 8197761. Signals are ordered 
according to decreasing amplitude. The values in braces are the formal 
errors of the frequencies according to \citet{mon99}. 
The formal errors of the amplitudes are $\pm 0.007$ mmag.}
%{\tiny{
\begin{tabular}{cccccccccccc}
\hline
ID & Frequency & Ampl. & Comment & ID & Frequency & Ampl. & Comment & ID & Frequency & Ampl. & Comment\\ 
 & [\cd] & [mmag] & &  & [\cd] & [mmag] & & & [\cd] & [mmag] &\\
\hline
$f_{1}$ & 1.0245220(5) & 5.715 & dominant & $f_{23}$ & 1.138304(6) & 0.461 & $f_{12}-f_{sp}$  & $f_{45}$ & 0.85572(1) & 0.268 & $f_{15}-f_{sp}$ \\
$f_{2}$ & 1.0983090(5) & 4.981 & dominant & $f_{24}$ & 2.074612(6) & 0.411 & & $f_{46}$ &  2.04904(1) & 0.247 & $2f_1$\\ 
$f_{3}$ & 0.7909331(6) & 4.204 & dominant & $f_{25}$ & 1.217526(7) & 0.398 & & $f_{47}$ & 0.31484(1) & 0.244\\
$f_{4}$ & 0.932967(2) & 1.669 & $f_{19}$-$f_{sp}$  & $f_{26}$ & 0.205307(7) & 0.397 & & $f_{48}$ & 0.65769(1) & 0.239\\
$f_{5}$ & 2.032936(2) & 1.301 & & $f_{27}$ & 1.746483(7) & 0.375 & & $f_{49}$ & 0.28160(1) & 0.230\\
$f_{6}$ & 1.096650(2) & 1.122 & $f_{2}-f_{sp}$  & $f_{28}$ & 1.022863(7) & 0.374 & $f_{1}-f_{sp}$ & $f_{50}$ & 0.81274(1) & 0.227\\
$f_{7}$ & 1.339370(3) & 0.998 & & $f_{29}$ & 0.399197(7) & 0.363 & & $f_{51}$ & 1.06970(1) & 0.220\\
$f_{8}$ & 1.099968(3) & 0.992 & $f_{2}+f_{sp}$  & $f_{30}$ & 1.955792(7) & 0.359 & & $f_{52}$ & 0.57238(1) & 0.215\\
$f_{9}$ & 0.619782(3) & 0.980 & & $f_{31}$ & 2.776998(8) & 0.354 & & $f_{53}$ & 0.10132(1) & 0.207 & transit freq.\\
$f_{10}$ & 2.196618(3) & 0.884 & $2f_2$ & $f_{32}$ & 2.122831(8) & 0.334 & $f_{1}+f_2$ & $f_{54}$ & 1.68808(1) & 0.206\\
$f_{11}$ & 0.811810(3) & 0.873 & & $f_{33}$ & 0.813469(8) & 0.330 & $f_{11}+f_{sp}$ & $f_{55}$ & 1.02618(1) & 0.204 & $f_1+f_{sp}$ \\
$f_{12}$ & 1.139963(3) & 0.864 & & $f_{34}$ & 1.959110(8) & 0.329 & $f_{30}+2f_{sp}$ & $f_{56}$ & 1.88924(1) & 0.200 & $f_2+f_3$\\
$f_{13}$ & 1.869165(3) & 0.785 & & $f_{35}$ & 0.992229(8) & 0.325 & & $f_{57}$ & 1.59242(1) & 0.192\\
$f_{14}$ & 0.163681(3) & 0.771 & & $f_{36}$ & 1.337711(8) & 0.321 & $f_{7}-f_{sp}$  & $f_{58}$ & 1.88192(1) & 0.186\\
$f_{15}$ & 0.857377(4) & 0.687 & & $f_{37}$ & 0.240950(8) & 0.315 & & $f_{59}$ & 1.75609(1) & 0.185\\
$f_{16}$ & 2.995046(4) & 0.679 & & $f_{38}$ & 1.815455(9) & 0.299 & $f_1+f_3$ & $f_{60}$ & 2.23833(2) & 0.178\\
$f_{17}$ & 1.538856(4) & 0.676 & & $f_{39}$ & 0.618123(9) & 0.295 & $f_9-f_{sp}$  & $f_{61}$ & 2.43772(2) & 0.176\\
$f_{18}$ & 1.792061(4) & 0.628 & & $f_{40}$ & 0.594144(9) & 0.291 & & $f_{62}$ & 1.85173(2) & 0.174\\
$f_{19}$ & 0.934627(4) & 0.605 & & $f_{41}$ & 0.470330(9) & 0.282 & & $f_{63}$ & 2.94869(2) & 0.173\\
$f_{20}$ & 0.936286(5) & 0.586 & $f_{19}+f_{sp}$  & $f_{42}$ & 1.09174(1) & 0.274 & & $f_{64}$ & 0.59325(2) & 0.172\\
$f_{21}$ & 0.859036(5) & 0.502 & $f_{15}+f_{sp}$  & $f_{43}$ & 1.14162(1) & 0.272 & $f_{12}+f_{sp}$  & $f_{65}$ & 0.65935(2) & 0.171 & $f_{48}+f_{sp}$ \\
$f_{22}$ & 0.404712(6) & 0.475 & & $f_{44}$ & 0.75510(1) & 0.268\\
\hline
\end{tabular}
%}}
\label{table:sc_frequencies}
\end{table*}

It is important to note that the transit frequency is detected in this analysis. This comes as a slight surprise because the transits had been removed from the data and therefore no leftovers from 
these features should be present in this data set. There are also several 
frequencies close to, but significantly different from, harmonics of the 
orbital period (in particular $f_{11}$, $f_{22}$, and $f_{25}$). The 
phase of the signal at the transit frequency excludes a reflection 
effect or ellipsoidal variability in case the orbital frequency had half 
this value. A possibility we cannot exclude is that this signal 
could correspond to a surface rotation period, most likely of the 
primary star. The first harmonic of the transit frequency does not 
exceed 0.1 mmag in amplitude.

Furthermore, 60 out of these 65 frequencies are explicable as first-order 
combinations of two of the other frequencies, within the temporal 
resolution of the data set ($\Delta T = 1470$\,d). The exceptions are 
$f_{16}$, $f_{31}$, $f_{40}$, $f_{42}$, and $f_{53}$ (the frequency of 
the transits). In Table~\ref{table:sc_frequencies} we have only 
explicitly marked the combinations generated by the three dominant 
oscillations.

Several of the dominant oscillations are part of multiplets (mostly 
triplets, sometimes doublets) equally spaced in frequency, and are often 
the centroid frequencies. To check whether the frequency spacing is the 
same for all these features, we computed a fit to the light curve under 
this assumption, as facilitated by \texttt{Period04}. We found this 
hypothesis to be consistent with the data, and can therefore determine the 
weighted mean frequency spacing of these multiplets:
\begin{center} 
$f_{sp} = 0.001659(15)$ \cd. 
\end{center}

After prewhitening these signals, some peaks in the 
higher-frequency part of the periodogram remain 
(Fig.~\ref{fig:power_spectrum}, lower panel). These are attributable to 
$\delta$ Scuti pulsations, and we summarize them in Table 
~\ref{table:dsct_freq}. Because the $\gamma$ Doradus domain in the 
periodogram could not be entirely prewhitened, formal error estimates 
for the frequencies and amplitudes of the higher-frequency signals are 
not meaningful. We therefore conservatively estimate the frequency error 
with one fourth of the Rayleigh frequency resolution \citep{krw08}, and 
the amplitude error with the local noise level in the periodogram. 
Contrary to the low-frequency domain, there is no evidence for regular 
frequency spacings among the $\delta$ Scuti pulsations.

\begin{table}
\centering
\caption{The $\delta$ Scuti pulsation frequencies detected in the 
\textit{Kepler} long cadence data of KIC 8197761. Signals are ordered 
according to decreasing amplitude. We estimate the errors with
$\pm 0.0002$\,\cd in frequency and $\pm$ 0.8\,$\mu$mag in amplitude
(see text for details).}
\begin{tabular}{ccc}
\hline
 ID & Frequency & Amplitude\\
 & (\cd) & ($\mu$mag) \\
\hline
$f_{A}$ &  17.1394 &   7.9 \\
$f_{B}$ &  16.3378 &   6.7 \\
$f_{C}$ &  18.6038 &   6.3 \\
$f_{D}$ &  16.3265 &   6.1 \\
$f_{E}$ &  20.9670 &   5.7 \\
$f_{F}$ &  16.6571 &   5.1 \\
$f_{G}$ &  21.8416 &   4.7 \\
$f_{H}$ &  17.1473 &   4.6 \\
$f_{J}$ &  17.2218 &   4.5 \\
$f_{K}$ &  19.6554 &   4.3 \\
$f_{L}$ &  21.8605 &   3.3 \\
\hline
\end{tabular}
\label{table:dsct_freq}
\end{table}
  
A fit composed of the frequencies determined from the earlier period analysis was then subtracted both from the entire SC and LC data set (now including the eclipses). The inclusion of frequency 
$f_{53}$ in this fit does not affect the analysis to follow as its 
parameters were determined from the light curve with the eclipses removed. 
% \begin{figure*}
%   \centering
%   \includegraphics[scale=0.36]{cleaned}
%   \caption[Long cadence \textit{Kepler} observations of KIC 8197761 with some pulsations removed]{Long cadence \textit{Kepler} observations of KIC 8197761 after removal of some part of $\gamma$~Doradus pulsations. Transit-like events are visible.}
%   \label{fig:cleaned}
% \end{figure*}
The cleaned SC light curve was used to obtain a rough estimate of the 
period of consecutive transits -- $9.87$ days. Then the Fourier transform 
of the residual LC data was computed (Fig.~\ref{fig:harmoniki}). As 
expected, a ``comb'' of harmonics of this frequency is present.
\begin{figure}
  \centering
  \includegraphics[scale=0.41]{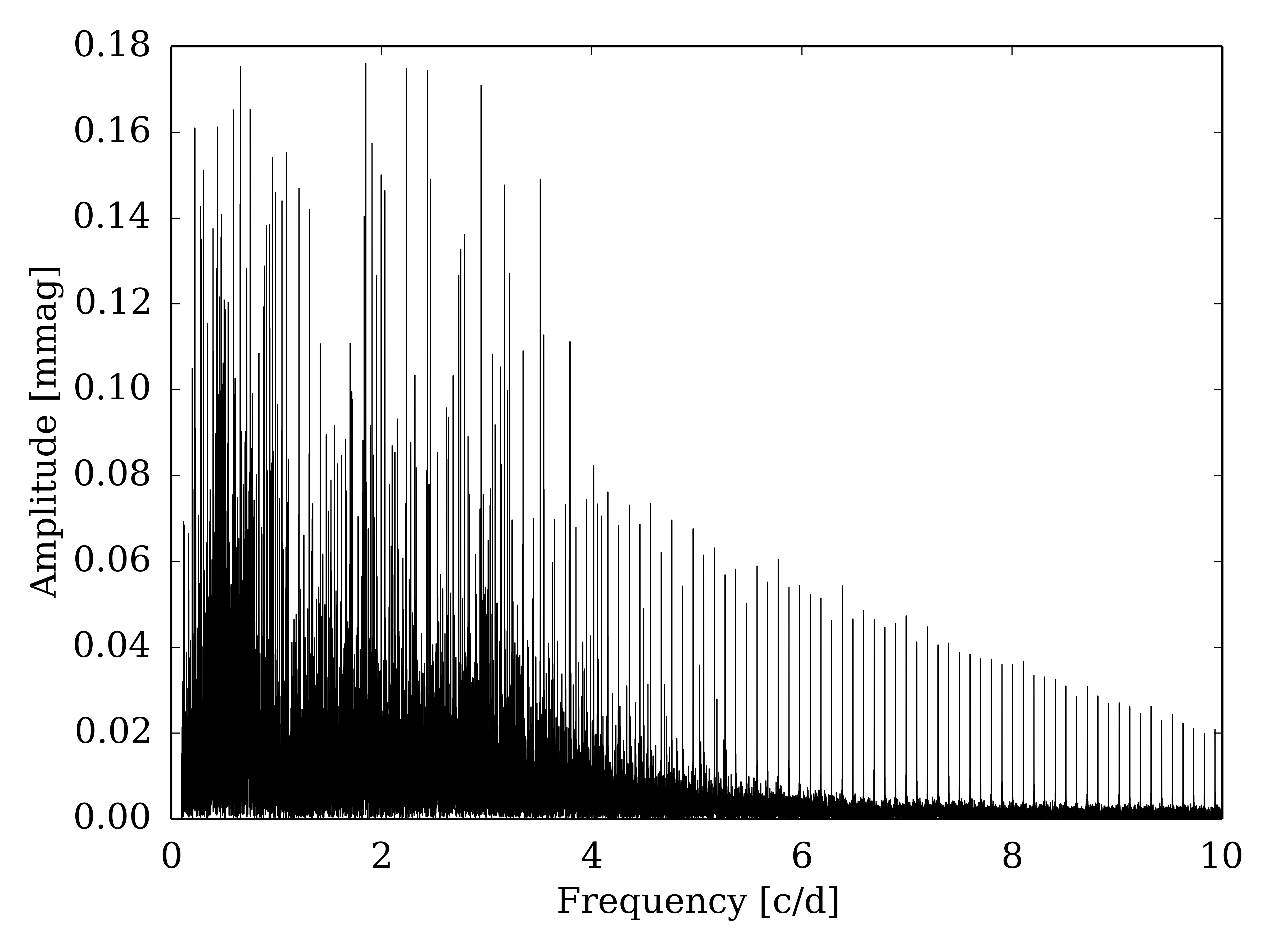}
  \caption[The Fourier calculation for the transit period (LC data)]{The Fourier 
calculation for the transit frequency (LC data) after prewhitening 65 
frequencies, showing a forest of harmonics. Those on the right-hand side 
of the plot were used to determine the final transit frequency.}
  \label{fig:harmoniki}
\end{figure}

The transit frequency and 59 of its harmonics (all clearly visible in 
the periodogram) were fitted to the data and their parameters optimised by 
fixing their frequencies to 
the exact harmonic multiples within \texttt{Period04}, allowing a 
precise determination of the corresponding period. The result is:
\begin{center}
$f_{trans} = 0.101330811(28)$ \cd $\Rightarrow P = 9.8686667(27)$ d,
\end{center}
which is in agreement with \citet{tenenbaum2012}.

The other observable which is easy to measure is the transit depth. The 
short cadence data were used to reveal the transit profile. Because we 
were not able to completely remove all of the pulsations from the light curve, the 
shape and depth of the transit was, unfortunately, affected. We fitted a linear function 
to the parts in the vicinity of the transit, but out-of-transit, to 
straighten the transit profiles. The resulting transit profile is shown 
in Fig.~\ref{fig:transit_zoom}. The change in flux was estimated as 
$\frac{\Delta F}{F} = 0.00629(29)$ and was applied to equation
\begin{equation}
\frac{\Delta F}{F} = {\left( \frac{R_p}{R_*} \right)}^2
\label{eq:depth}
\end{equation}
with a~$R_*~=~1.717~R_{\sun}$ from \citet{huber2014}. Hence the 
companion's minimum radius is found to be:
\begin{center}
$R_{companion} \geq 0.136 \ R_{\sun}$% = 1.35 R_J$. %= 11.56_{-3.49} R_{\oplus}$.
\end{center}

%----------------
\begin{figure}
  \centering
  \includegraphics[scale=0.41]{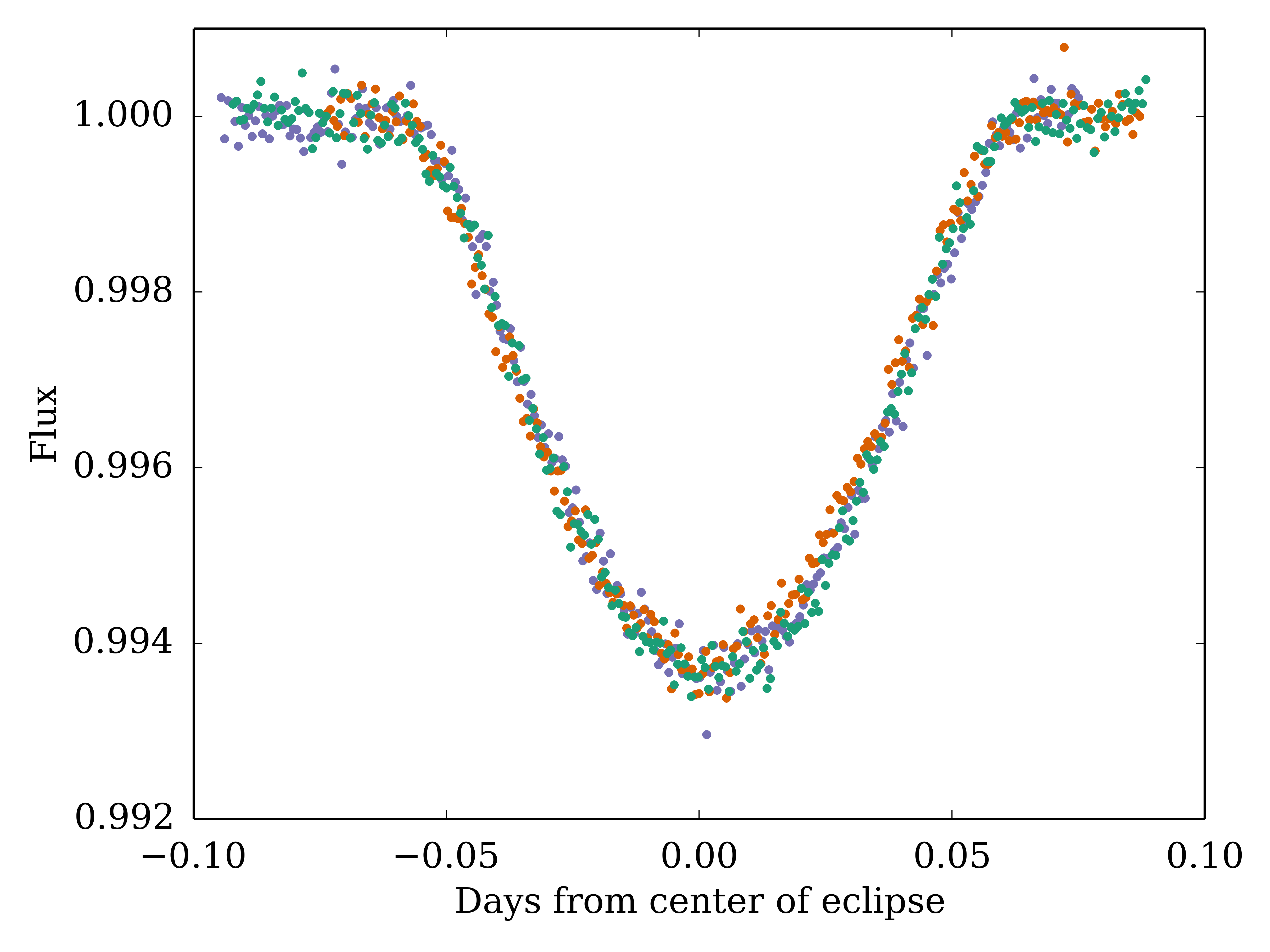}
  \caption[The transit profile in zoom]{The possible transit profile 
from short cadence data in zoom. Blue points: the first transit in the 
data, orange points: the second transit in the data, shifted by the 
transit period, green points: the third transit in the 
data, shifted by twice the transit period (colour plots available in the online version of this article).}% The duration of the transit is 8.37 h.}
  \label{fig:transit_zoom}
\end{figure}
One can notice that the shape of the transit shows a V-like profile rather than typical (for planetary eclipses)
U-like with a nearly flat minimum, perhaps indicating a grazing eclipse and/or that the companion has a radius comparable to that of the primary (as indicated by the minimum companion radius derived above).  By phasing the light 
curve with respect to the transit period there is no appreciable
secondary eclipse/transit. 
%------------------------------------------------------------------------------------------------------------------------
\subsection{Asteroseismology}
\label{sec:asteroseismology}
In Section~\ref{sec:per}, we reported the detection of an equal frequency splitting 
within several of the pulsational signals detected. As we classified KIC 
8197761 as a $\gamma$ Doradus pulsator, these signals would be due to 
gravity-mode pulsations. It follows that they are likely modes of the same 
spherical degree $l$ split by rotation into $2l+1$ components of azimuthal 
order $m$ and different radial overtone $k$. Asymptotic pulsation theory 
(e.g., \citealt{tassoul1980}) predicts that such modes with consecutive radial 
overtones should be equally spaced in period.

Equal spacings of signals within data sets can be revealed by computing 
their spectral window (see \citealt{handler1997} for a detailed description 
of the method). We therefore performed a corresponding search for regular 
spacings within the 65 frequencies in Table~\ref{table:sc_frequencies} and 
show its result in Fig.~\ref{fig:kicpspw}.

\begin{figure}
  \centering
  \includegraphics[scale=0.41]{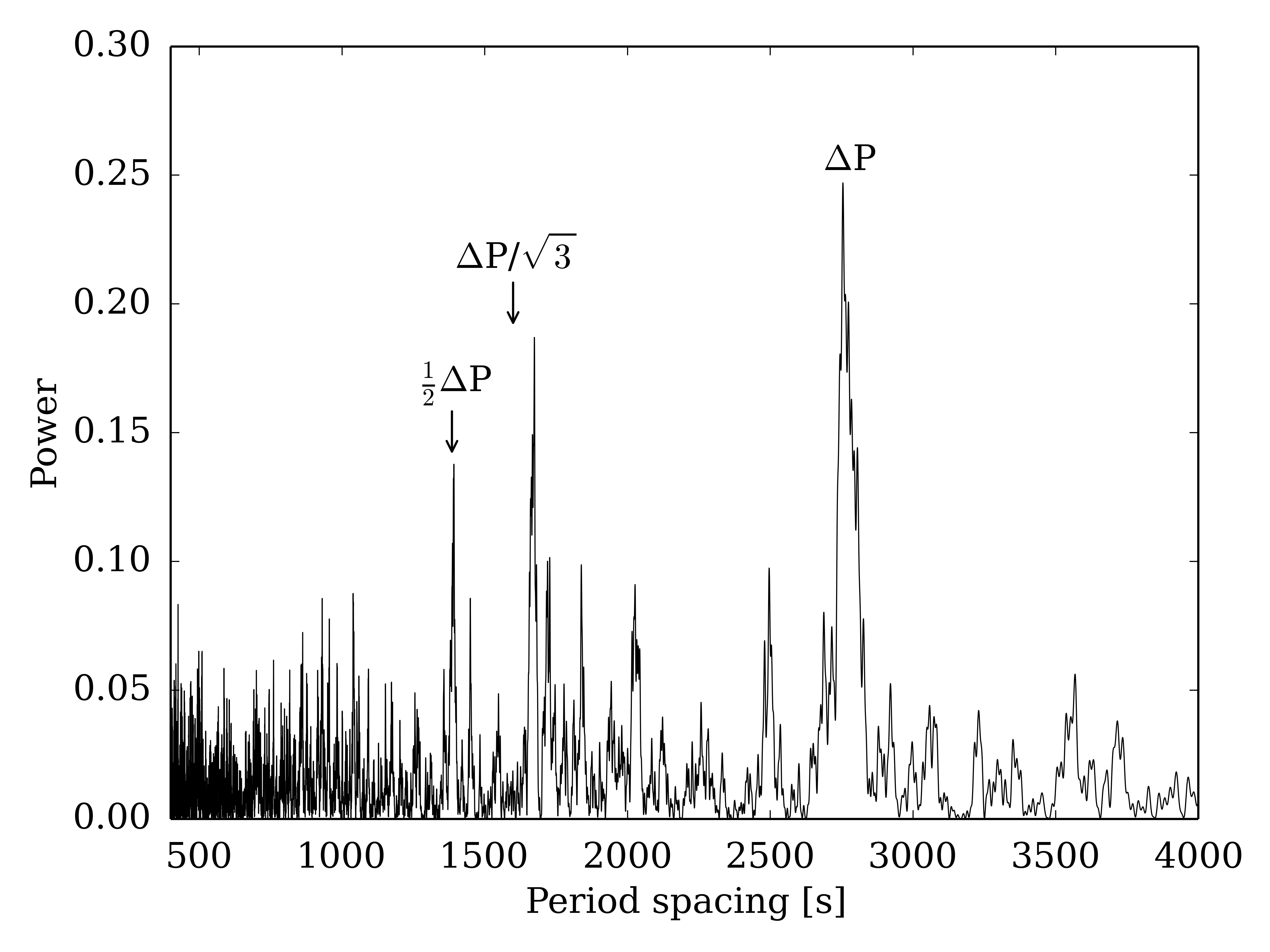}
  \caption[The spectral window of the period spectrum]{The spectral window 
of the period spectrum of KIC 8197761. A mean period spacing around 2770 s 
and its first harmonic are indicated. Another strong signal near to,
but significantly different from, $2770/\sqrt{3}$ s is also present.}
  \label{fig:kicpspw}
\end{figure}

This analysis suggests the presence of a mean period spacing of $2770 
\pm 40$ s between the pulsation modes. A second possibility for a mean 
period spacing of $1670\pm20$ s is also indicated. This is interesting, 
as the mean period spacings of high-order gravity modes such as excited 
in $\gamma$ Doradus stars should relate as $\Delta P_{l=1}/\Delta 
P_{l=2}=\sqrt{3}$ \citep{unno1989}. However, the ratio of these two 
potential mean period spacings is $1.66\pm0.04$, almost $2\sigma$ from 
the asymptotic value. As our attempts to reproduce such an $l=2$ 
period spacing with models with a fixed $l=1$ spacing of $2770 \pm 40$ s 
failed (mode trapping could cause deviations from the theoretically 
expected $\Delta P_{l=1}/\Delta P_{l=2}$ period ratio), we consider this 
peak accidental.

\begin{figure}   
  \centering
  \includegraphics[scale=0.47]{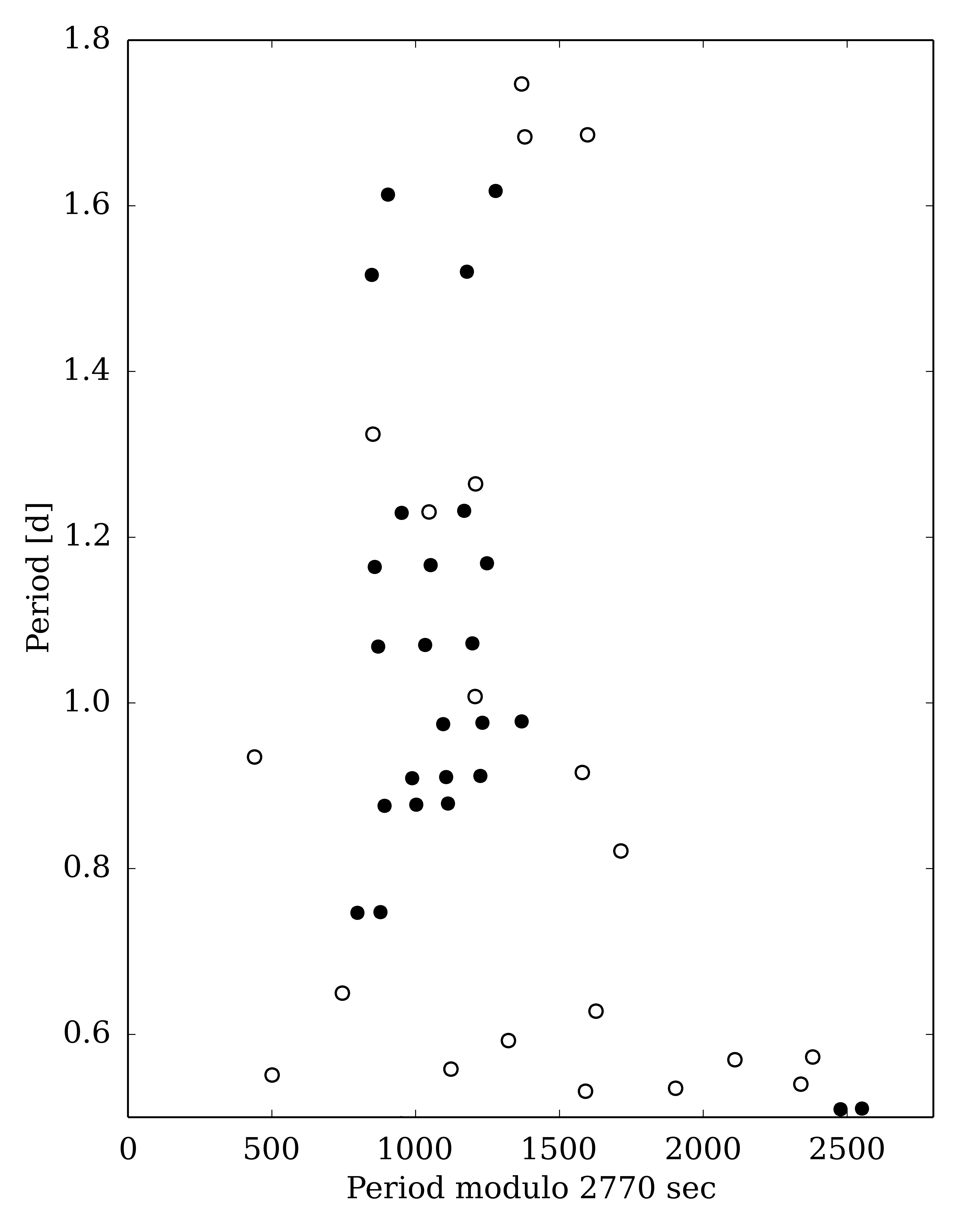}
  \caption[Echelle Diagram]{Echelle Diagram of the pulsation 
periods of KIC 8197761. Full circles correspond to signals which are 
components of multiplets with equal frequency splittings, whereas the open 
circles are the remaining signals from Table~\ref{table:sc_frequencies}.}
  \label{fig:kicechp}
\end{figure}

An Echelle Diagram with respect to the 
2770 s mean period spacing is shown in Fig.~\ref{fig:kicechp}. The 
signals that are components of multiplets with equal frequency splittings 
roughly fall onto a vertical sequence. An exception are the two 
shortest-period of these signals, $f_{30}$ and $f_{34}=f_{30}+2f_{sp}$. 
However, as $f_{1}+f_{4}=f_{30}+f_{sp}$, it is likely that these two 
signals are in fact combination frequencies. Therefore, we conclude 
that the equally-split frequency multiplet structures with periods between 
$0.7-1.7$ d are indeed due to $g$ modes of the same $l$.

There are three main reasons to conclude that these signals 
correspond to $l=1$ modes. Firstly, the multiplets have a maximum number of 
three members, which is the maximum number of $m$ components possible for 
$l=1$. If the spherical degree was higher, it would be improbable to see 
only frequency doublets and triplets. Secondly, the mean $l=1$ period 
spacing of Zero-Age Main Sequence models of $\gamma$ Doradus stars is 
about 3000 s (see \citealt{miglio2008}). The mean period spacings of modes 
of higher $l$ would be at least by $\sqrt{3}$ smaller. In other words, if 
2770 s was the mean period spacing of modes $l>1$, the corresponding $l=1$ 
period spacing would be considerably in excess of 3000 s, which is 
incompatible with models in the parameter space of $\gamma$ Doradus stars. 
Thirdly, geometrical cancellation over the visible stellar disk 
\citep{dziembowski1977} favours the observation of $l=1$ modes in 
comparison to higher $l$.

Returning to Fig.~\ref{fig:kicechp}, the $l=1$ multiplets do not lie in a perfect vertical ridge in the Echelle Diagram; a ''wavy`` structure is 
superposed. This is an effect of mode trapping, caused by the sharper 
density gradients in the stellar interior as the object evolves, and can 
be taken advantage of to constrain its evolutionary state (see, e.g., 
\citet{saio2015} and \citet{reeth2015} for examples).

We conclude that the frequency multiplets in 
Table~\ref{table:sc_frequencies} are due to rotationally split $l=1$ g 
modes of high radial order. With the rotational splitting of $f_{sp} = 
0.001659(15)$ \cd, and in the asymptotic limit therefore
\begin{displaymath}
P_{rot}=\frac{1-[l(l+1)]^{-1}}{f_{sp}},
\end{displaymath}
which results in $P_{rot}=301\pm3$ d. We note that an initial attempt of 
more detailed asteroseismic modelling of the pulsation spectrum of KIC 
8197761 did, aside from confirming the discussion above, not bear fruit 
(H. Saio, private communication). 

%------------------------------------------------------------------------------------------------------------------------- 
\subsection{The Light Travel Time Effect} 

We now return to the question 
of the nature of the companion of the $\gamma$ Doradus pulsator. Having 
only a radius estimate (or a lower limit) for the companion available, 
it is not possible to rule out a stellar companion; an estimation of the 
secondary's mass is crucial. There are many methods that can be used to discover planets and/or 
to measure their masses, like the radial velocity method, gravitational 
microlensing, the timing method, variability of the star in phase with 
the planetary orbit due to beaming, ellipsoidal variability, and 
reflection (e.g. \citealt{faigler2011}, \citealt{szabo2012}). In the case of 
the LTTE method, some observables modulated by the presence of a 
companion can be used. For a pulsating star, these would be the 
frequencies of pulsations which are stable over time.

The idea of using the timing method for pulsating stars is not new, there 
are many earlier applications, e.g. \citet{breger1987}. Also  
recently \citet{murphy2014} showed that using phase modulation of 
$\delta$ Scuti pulsations it is possible to create radial velocity curves 
and discover binary companions (also determining the system parameters).  The orbital motion (due to the presence of a companion star or planet) of a pulsating star acts to alter the arrival times of observed pulsations, producing phase and frequency modulations in the observed light curve. Measuring these modulations can allow the mass of the companion to be estimated.

The amplitude of this effect, for the case of inclination near 
$90^{\circ}$ (i.e. a system with transits) is:
\begin{equation}
t_{LTTE} = \frac{m_2}{M_* + m_2} \cdot \frac{a}{c},
\label{eq:ltte}
\end{equation}
while for a non-transiting planets $m_2$ is replaced by $m_2\sin i$ in 
Eq.~\ref{eq:ltte}.

%-------------------------------------------------------------------------------------------------------------------------
%\subsection{Constraints on the companion's mass - three scenarios}
%\label{sec:scen}
Having the transit period determined very accurately, the next step was to 
remove eclipses/transits from the light curve with pulsations not subtracted.
The times of these events were determined from the 
pulsation-removed light curve phased at the transit frequency.
Then, a simultaneous fit with 340 frequencies, except for the pulsation under consideration, was subtracted from the light curve. These are more frequencies than listed in Table 3. The removal of the additional low-amplitude signals merely served the purpose to decrease the variance in the residual light curve. As we are not convinced that all of these are real, we prefer not to list them here. After that, the light 
curve was divided into parts that were integer fractions (20) of the 
transit period and these small pieces representing the same orbital 
phase range were put together again.

The phases of the pulsations under consideration 
for determining the size of the LTTE were then determined one by one in 
these 20 individual subsets with respect to the transit phase. If a 
significant variation of those phases with the transit period can be 
found, their amplitudes (converted from phase to time given the known 
pulsation periods) give the size of the orbital LTTE. Given the parameters of the primary star, the light travel time change can then be used to place a lower limit on the companion mass. If no significant 
orbital light time effect is found, only an upper limit for it and 
therefore an upper mass limit for the companion can be estimated. The 
hope was that this would have been sufficient to decide whether the 
companion is a star or a planet.

To determine the LTTE, and given the complicated pulsation spectrum 
of KIC 8197761, it may be advantageous to include a large number of 
stellar pulsation frequencies. On the other hand, the phase 
determinations for weaker oscillations may be distorted by the presence 
of residual variability in the small data subsets to be fitted. The 
combined influence of these two factors on the precision of the LTTE 
measurement is difficult to foresee. Therefore, we experimented by 
determining the LTTE from unweighted or weighted averages of the 
individual results from up to 60 pulsation frequencies. Weights were 
assigned according to the product of the amplitude and frequency of a 
given signal: the higher its amplitude, the more precise its phase 
measurement, and the higher the frequency, the larger the phase change 
in units of time \citep[cf.][]{mon99}.

%-------------------------------------------------------------------------------------------------------------------------
%\subsubsection{The first scenario}

There are three possible scenarios for the cause of the 
transits/eclipses in the system. These would be a) two stars of 
equal mass, b) two stars of different mass, or c) a star and a planet. 
These will be discussed in the remaining part of this section.
The hypothesis of two stars of equal mass in this system stems from the 
lack of a secondary eclipse/transit. In this scenario (i), the true orbital 
period is twice the period determined from the eclipses/transits and 
therefore primary and secondary eclipse are indistinguishable.

The expected LTTE in this scenario, assuming two stars of $1.384$ 
M$_{\odot}$, is easy to calculate using Kepler's third law and 
amounts to $t_{LTTE,theoretical} = 49$ s.

%-------------------------------------------------------------------------------------------------------------------------
%\subsubsection{The second and third scenario}
The other two scenarios assume that the period determined in 
Section~\ref{sec:per} is the true orbital period. In that case, for 
the LTTE is limited to $t_{LTTE,theoretical} < 32$ s, while for substellar secondaries this is reduced to $t_{LTTE,theoretical} < 2.8$ s.

The best observational limit on the LTTE we achieved with our 
trials was $t_{LTTE,obs}=4\pm20$ s. This was obtained with weights 
assigned as $\sigma^{-2}$ and including 30 frequencies or more. As a 
result, the $1\sigma$ upper level we can put on a possible companion's 
mass is $m_2<0.94$\,M$_{\odot}$. This is a rather weak constraint, but 
at least sufficient to provide an argument against the scenario of two 
equal-mass stars in a 20-d orbit.

What remains to be concluded is that with the long periods and low 
amplitudes of the pulsations in $\gamma$ Doradus stars, detection of the LTTE caused by a substellar 
companion seems impossible, except in special circumstances (many 
oscillation modes of high amplitude, additional presence of 
$\delta$~Scuti pulsations of sufficient amplitude, or long orbital 
period). \cite{compton2016} performed a number of simulations reaching the same conclusion.

%-------------------------------------------------------------------------------------------------------------------------

\subsection{Spectroscopy}
To constrain the nature of the companion to KIC 8197761 further, we 
obtained a series of 13 spectra over 1.5 times the photometric transit 
period. The observations were performed in late June/early July 2015 using the 
High Efficiency and Resolution Mercator Echelle Spectrograph 
(\texttt{HERMES}, \citealt{raskin2011}) attached to the 1.2-m Mercator 
Telescope, located at the Roque de los Muchachos Observatory on La 
Palma, Spain. The spectra covered the wavelength range from 
$3770-9000$\AA\, with a resolution of $R=85000$. Table~\ref{table:spectra} shows the journal of the spectroscopic observations.

\begin{table*}
\centering
\caption{Log of spectroscopic observations of KIC 8197761 with the 1.2-m Mercator Telescope. The error is the rms of the residuals of the sinusoidal fit to the individual radial velocities.}
%{\tiny{
\begin{tabular}{ccc}
\hline
Obs. time (JD) & Exp. time (s) & Deviation from systemic velocity (km s$^{-1}$) \\ 
\hline
2457192.57352 &  900 & -2.09(47) \\ 
2457194.63324 &  900 & -19.31(47) \\
2457195.65493 &  900 & -17.45(47) \\
2457196.60316 &  900 & -10.52(47) \\
2457197.64462 &  800 & 2.28(47) \\
2457198.66634 &  900 & 14.94(47) \\
2457199.69848 &  900 & 19.45(47) \\
2457200.68586 &  900 & 17.41(47) \\
2457201.66651 &  900 & 8.37(47) \\
2457202.64345 &  900 & -3.36(47) \\
2457203.59466 &  900 & -13.54(47) \\
2457205.59388 &  900 & -17.73(47) \\
2457207.56990 &  900 & 3.27(47) \\
\hline
\end{tabular}
%}}
\label{table:spectra}
\end{table*}

All spectra were reduced using the dedicated \texttt{HERMES} 
pipeline. Radial velocity analysis was performed using the 
\texttt{molly}{\footnote{http://deneb.astro.warwick.ac.uk/phsaap/software/molly/html/INDEX.html}} software package. 
Barycentric corrections were applied to all spectra, before cross-correlating each against the first spectrum in the series (dividing each spectrum into four pieces of equal wavelength range to aid the process computationally).  The first spectrum was arbitrarily chosen as the template spectrum, which was deemed appropriate given that each spectrum had a similarly high signal-to-noise.  It is important to note that by choosing one of the observed spectra as the template, we are unable to determine the systemic velocity of the binary rather only the variability (as the template spectrum arbitrarily sets the zero against which all other velocities are measured).
The resulting 
measurements are a weighted average of values obtained for each spectrum 
chunk, where weights were typical $1/\sigma^2$ (the standard deviation 
of the set of velocity measurements from all chunks of the spectrum). The ''radial velocity`` 
curve obtained this way was then fit with a sine function with fixed 
period - the orbital period of $P = 9.8686667(27)$ d (Fig.~\ref{fig:rv}). 
It is clear that the photometric transit period is indeed the orbital 
period, and that the observed eclipses are only primary 
eclipses/transits. The RV curve is well fit with a sine function of 
amplitude $K_1 = 19.75 \pm 0.32$~km/s. \citet{sant2016} derived a value
consistent with ours from only two RV measurements. We checked the presence of a possible small orbital eccentricity in our radial velocity data, and determined a best-fitting value of $e = 0.02 \pm 0.02$. We consider this result as not significant at the time being and stay with the assumption of a circular orbit.

\begin{figure}
  \centering
  \includegraphics[scale=0.41]{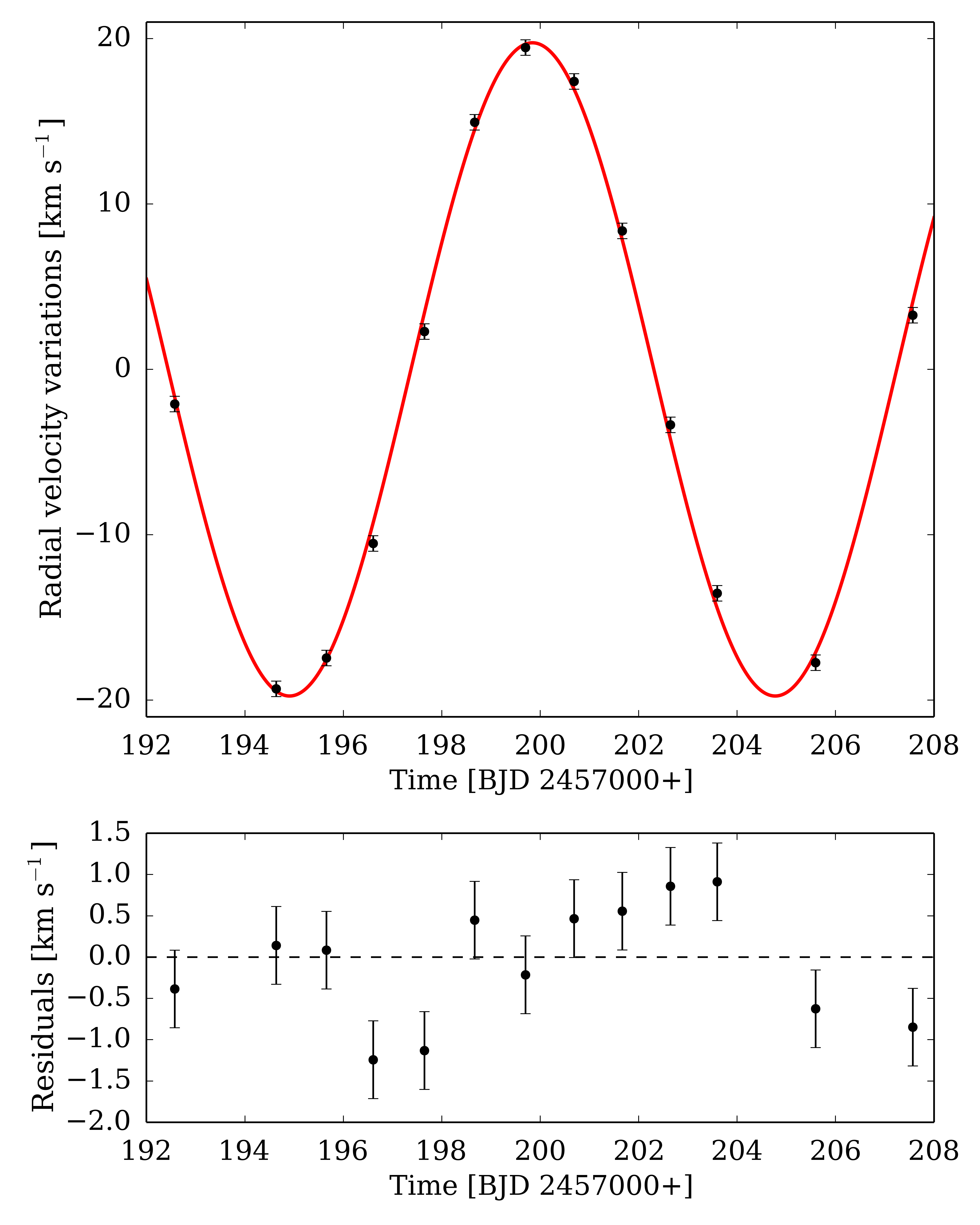}
  \caption{Upper panel: The radial velocity measurements of KIC 8197761 with fit of a sine function. This is radial velocity variability shifted so that the systemic velocity is 0. Lower panel: Residuals to the circular orbit fit.}
  \label{fig:rv}
\end{figure}

The amplitude of the derived radial velocity curve, when combined with the mass estimate of the primary (and assuming an orbital inclination of 90\degr{}) results in a lower limit secondary mass of $m_2 = 0.28$ M$_{\sun}$.  This mass would correspond to a mid-M dwarf, the typical radius of which would be $\sim$0.3R$_{\sun}$ (e.g., \citealt{feid2012}), accounting for this the eclipse duration implies an orbital inclination $i\gtrsim$85\degr{}.  As such, the companion is not of planetary mass.

%{\bf The mass function of the secondary, $f(M)=P\frac{P K_1{^3}}{2 \pi G}$, is 0.00787~M$_{\sun}$,
%accounting for the mass estimate of the primary and assuming an inclination of 90\degr
%results in a minimum secondary mass of $m_2 = 0.28$ M$_{\sun}$. Such a mass would 
%correspond to a mid-M dwarf, the typical radius of which is around 0.3~R$_{\sun}$ (e.g., \citealt{feid2012}).}
%\sout{The secondary mass function using $m_2=\frac{P K_1{^3}}{2 \pi G}$ is 
%0.00787~M$_{\sun}$. Given the mass estimate of the primary star 
%(Table~\ref{table:parameters}), an orbital inclination of 90\degr results in a minimum 
%secondary mass of $m_2 = 0.28$ M$_{\sun}$, corresponding to a mid-M dwarf. 
%The typical radius for an object of this mass is around 0.3 R$_{\sun}$ 
%(e.g., reference). Using this estimate and Kepler's third law, we 
%determine a minimum orbital inclination of $i=85\degr$. Consequently, for 
%any reasonable primary mass, the companion is not of planetary mass.

We also determined the projected rotational velocity of KIC 8197761 from 
our spectra. To this end, the individual spectra were shifted by the radial 
velocity offsets in Table~\ref{table:spectra} and then averaged. A synthetic spectrum for the star was then computed using the 
program \texttt{SPECTRUM} 
\citep{gray1994}\footnote{http://www.appstate.edu/$\sim$grayro/spectrum/spectrum.html} 
and an ATLAS9 model atmosphere \citep{castelli2004} for $T_{\rm 
eff}=7250$ K, log $g$=4.0, $[M/H]=0.0$, and a microturbulent velocity of 
2 km/s. We broadened the synthetic spectrum with trial projected 
rotational velocities and the instrumental profile as determined from 
the telluric lines in the \texttt{HERMES} spectra, and compared it with 
the observed average. The best fit was achieved with $v \sin i=9\pm1$ km/s. 
This value can still be affected by macroturbulence and pulsational 
line broadening. To estimate the latter, we adapt the method and result 
by \mbox{\citet{murp16a}} who computed the expected pulsational line broadening 
for another $\gamma$ Doradus star, KIC 7661054. These authors computed a 
line broadening of $\approx 2$ km/s as a result of the star's pulsation. 
As the photometric peak-to-peak amplitude of KIC 8197761 is some 30\% 
larger than that of KIC 7661054, we estimate the pulsational line 
broadening with $\approx 2.6$ km/s for our star. Keeping in mind that 
all line broadening mechanisms add up as the sum of squares, pulsational 
line broadening would not alter our $v \sin i$ determination outside the 
quoted errors; the same comment applies to macroturbulence up to 4 km/s. 
We therefore believe that our $v \sin i$ measurement corresponds to the real value within the errors.

Given the stellar radius in Table~\ref{table:parameters}, such a low rorational broadening would imply a rotation period $P_{\rm rot} < 10.9$~d, markedly 
different from the rotation periods of the star photometrically determined 
by \cite{niel2013} and \cite{rein2013}. This discrepancy is easily 
explained: both authors identified one of the stellar pulsation periods 
($f_1$ and $f_3$, respectively) as a rotational signal. More 
interestingly, the (surface) rotation period determined here is very 
close to the system's orbital period, but much shorter than the 
rotation period determined from the frequency splittings of the g-mode 
pulsations. It is therefore quite possible that the rotation of the 
envelope of KIC 8197761 is synchronised with the orbit (consistent with the circularised orbit found from the radial velocity curve), whereas the 
stellar core rotates 30 times slower than that.

%-------------------------------------------------------------------------------------------------------------------------

\section{Summary and conclusions}
The goal of this work was to search for exoplanets around pulsating 
stars. Given the large number of objects observed by 
\textit{Kepler} and the quality of the associated data, it is only now possible to 
perform such a search.

The stars for the search were taken from the MAST archive within a given 
effective temperature range, in order to isolate probable pulsators. That resulted in 7546 individual FITS files for a total of 2292 stars. The whole sample contained stars of different 
kinds of variability, e.g. $\delta$ Scuti, $\gamma$ Doradus, RR Lyrae, 
eclipsing binaries and more. The light curves were cleaned of isolated outlying points, and then prewhitened of signals found by Fourier analysis. The residual light curves were inspected for the 
presence of transits three times following progressively stricter criteria, resulting in a list of transit candidates consisting of 42 stars. Within the transit candidates 
there is only one, KIC 8197761, which shows exactly what was of 
interest in this work -- transit-like events hidden in pulsations. A 
second star, KIC 5613330, showed periodic transit-like events but no 
intrinsic, pulsational variability.

For KIC 8197761, a careful examination of the pulsational signals revealed the presence of an equal 
frequency splitting of $f_{sp} = 0.001659(15)$ \cd within some modes of 
oscillation. These modes were found to be rotationally split $l=1$ g 
modes of high radial order. The rotation period obtained using $f_{sp}$ 
and in asymptotic limit is $P_{rot}=301\pm3$ d. Such a 
period is very long for a star as hot as a $\gamma$ Doradus star, however, it is such a slow rotation that favours the detection of 
rotational splitting signals. An attempt to perform asteroseismic 
modelling (H. Saio, private communication) did not bear fruit. The 
eleven low-amplitude $\delta$ Scuti pulsations detected did not allow to 
obtain additional seismic information.

Returning to the eclipses/transits, a V-like light curve shape is 
present, suggesting that this may be grazing event. In such an 
ambiguous case a mass estimation is crucial, but there were no such 
constraints in the literature. Therefore, we attempted to estimate the mass 
using the LTTE, by utilising pulsation frequencies stable in time to 
measure orbitally induced phase variations of these pulsations.

Because there is no indication of the secondary eclipse/transit in the 
light curve, three possible scenarios were tested: two stars of equal 
mass, two stars of different mass and a~star and a planet. 
Unfortunately, given the errors of the determinations of the LTTE, 
only the first of these scenarios could be rejected. 
The rather long periods of $\gamma$ Doradus pulsations and their low 
amplitudes therefore make it difficult to detect planetary, and even 
low-mass stellar, companions around them using the LTTE.

As a final step we performed a spectroscopic analysis of KIC 8197761. Our 
13 spectra obtained with \texttt{HERMES} on the 1.2-m Mercator Telescope 
show radial velocity variations with an amplitude of $19.75 \pm 0.32$ 
km/s, and allowed the determination of $v \sin i=9\pm1$ km/s. The 
latter suggests that the surface rotation of the star is synchronized with 
the orbital period, although the stellar core rotates much slower. The 
estimated companion mass is $\approx 0.28$ M$_{\sun}$. Based on this 
result we reject the hypothesis of the companion being an exoplanet.

The combination of asteroseismology and exoplanetary research 
results in multiple synergies.
All the space missions designed to search for planets provide also 
excellent data to use for asteroseismic studies. Asteroseismology plays 
a key role especially in characterisation of the host stars. For example 
for solar-like stars the radius, the mass, and the age can be estimated 
with remarkable precision (of the order of a few percent, e.g. 
\citealt{met2010}). In the future, this may serve as a key test of our 
stellar models and our understanding in stellar physics. Another 
interesting question is the possible interaction(s) between the exoplanets and host 
stars, which may play a role in some of the mysterious pulsational behaviour 
observed in some stars (see \citealt{wri2011} for an example). Recently, 
asteroseismic methods began to serve as another exoplanet detection 
method by measuring the delay in pulsation time arrivals, and the first 
such planet around an A-type star has been discovered (\citealt{murp16b}). 
With the current and upcoming missions such as K2, TESS, or PLATO, this 
demonstrates how asteroseismology may aid to find and characterise exoplanets 
in the near future. These examples highlight that exoplanetary science 
would be less successful without asteroseismology and cooperation 
between both is required for a deep understanding of both stars and 
planets.

%-------------------------------------------------------------------------------------------------------------------------
\section*{Acknowledgments}

This paper includes data collected by the \textit{Kepler} mission. 
Funding for the \textit{Kepler} mission is provided by the NASA Science 
Mission directorate. All of the data presented in this paper were 
obtained from the Mikulski Archive for Space Telescopes (MAST). STScI is 
operated by the Association of Universities for Research in Astronomy, 
Inc., under NASA contract NAS5-26555. Support for MAST for non-HST data 
is provided by the NASA Office of Space Science via grant NNX13AC07G and 
by other grants and contracts. Based on observations made with the 
Mercator Telescope, operated on the island of La Palma by the Flemmish 
Community, at the Spanish Observatorio del Roque de los Muchachos of the 
Instituto de Astrof\'isica de Canarias. Based on observations obtained 
with the HERMES spectrograph, which is supported by the Research 
Foundation - Flanders (FWO), Belgium, the Research Council of KU Leuven, 
Belgium, the Fonds National de la Recherche Scientifique (F.R.S.-FNRS), 
Belgium, the Royal Observatory of Belgium, the Observatoire de Gen\'eve, 
Switzerland and the Th\"uringer Landessternwarte Tautenburg, Germany. 
PIP is a Postdoctoral Fellow of the The Research Foundation - Flanders 
(FWO), Belgium. We thank Tom Marsh for the use of \texttt{molly}. This 
work was partially supported by the Polish NCN grants 
2011/01/B/ST9/05448 and 2015/18/A/ST9/00578. We thank Hideyuki Saio, 
Michael Endl, Simon Murphy and Timothy van Reeth for helpful 
discussions, as well as Katrien Kolenberg for assistance with organizing 
the Mercator observations. An anonymous referee has provided many 
constructive comments that helped to improve this paper.}

%%%%%%%%%%%%%%%%%%%%%%%%%%%%%%%%%%%%%%%%%%%%%%%%%%



%\bibliographystyle{mnras}


%%%%%%%%%%%%%%%%%%%%%%%%%%%%%%%%%%%%%%%%%%%%%%%%%%

%%%%%%%%%%%%%%%%% APPENDICES %%%%%%%%%%%%%%%%%%%%%

%%%%%%%%%%%%%%%%%%%%%%%%%%%%%%%%%%%%%%%%%%%%%%%%%%

% Don't change these lines
\bsp	% typesetting comment
\label{lastpage}
\end{document}